# High-speed hyperspectral 3D ghost imaging LiDAR

Jing Hu[1]†, Wending Huang[1]†, Tianjian Lv[1]†, Ming Yan[1,2]*, Zhaoyang Wen[1,2], Zijian Wang[1], Yuan Chen[1], Zhuoren Wan[1], Mei Yang[1], and Heping Zeng[1,2,3]

[1]State Key Laboratory of Precision Spectroscopy, and Hainan Institute, East China Normal University, Shanghai, China

[2]Chongqing Key Laboratory of Precision Optics, Chongqing Institute of East China Normal University, Chongqing 401120, China

[3]Jinan Institute of Quantum Technology, Jinan, Shandong 250101, China

†These authors contribute equally.

* Corresponding author: *myan@lps.ecnu.edu.cn*

**Abstract:**

Light detection and ranging (LiDAR) is widely used in autonomous systems and industrial metrology; however, the simultaneous acquisition of three-dimensional (3D) structure and broadband spectral information remains challenging, as conventional hyperspectral LiDAR relies on wavelength-scanning or spectrometer-based detection that limits speed. Here, we demonstrate a hyperspectral 3D ghost imaging LiDAR that eliminates these bottlenecks. By combining a stochastic broadband laser with single-pixel detection, and integrating spatiotemporal encoding with spectral ghost imaging in a time-of-flight framework, the system enables pulse-resolved recovery of spatial and spectral information. Consequently, we achieve a line-scanning rate of 60.5 MHz (point rate 1.8 GHz) and a ranging precision of 0.02 mm within a 10 µs integration time. Each voxel contains a 1.4 nm resolution spectrum over 1100-1250 nm, enabling simultaneous 3D imaging and chemical identification. This approach provides a route to high-speed hyperspectral LiDAR for environmental monitoring, precision agriculture, and industrial inspection.

# Introduction

LiDAR has emerged as a foundational technology for remote sensing, enabling precise three-dimensional (3D) mapping and motion analysis in applications ranging from autonomous navigation to industrial inspection [1,2]. Recent advances have significantly expanded its capabilities through the integration of optical frequency combs [3-9], together with advanced beam-steering strategies such as optical phased arrays [10], acousto-optic deflectors [11], focal-plane switch arrays [12], and spatio-spectral encoding [13]. These innovations have driven substantial improvements in acquisition speed, enabling high-throughput 3D ranging and velocity sensing [14-17]. In contrast, LiDAR systems that simultaneously acquire co-registered spatial and spectral information—thereby enabling hyperspectral 3D imaging—remain comparatively underdeveloped, despite their importance for applications such as precision agriculture [18], environmental monitoring [19], and hazardous material detection [20], where both structural and chemical specificity are required.

Conventional hyperspectral LiDAR systems combine time-of-flight (ToF) ranging with spectrally resolved detection, typically implemented via wavelength scanning [21] or dispersive elements in the receiver [22]. However, their reliance on mechanical scanning and sequential spectral acquisition limits imaging speed, while the use of spectrometers or detector arrays increases system complexity and restricts applicability to predominantly static scenes [1, 2, 23].

Computational imaging offers a compelling alternative paradigm that eliminates the need for spatially resolving detectors [24]. While computational hyperspectral imaging has achieved video-rate acquisition, it typically lacks depth resolution, and its reliance on digital micromirror devices (DMDs) limits further improvements in speed [25]. Ghost imaging, a subset of computational imaging, reconstructs object information from correlations between structured illumination and bucket detector signals [26]. Originally developed in the spatial domain, ghost imaging has been extended to the time domain for ultrafast signal processing [27, 28], and further to the spectral domain by encoding wavelength information into stochastic spectral patterns. More recently, the integration of stochastic light sources [29-31] with time-stretch dispersive Fourier transformation (TS-DFT) has enabled ultrafast, pulse-by-pulse spectral encoding and reconstruction. In parallel, TS-DFT-based LiDAR systems have been demonstrated [32], in which depth information is extracted via ToF measurements while lateral spatial information is mapped onto the spectral axis. Despite their high imaging speeds, these systems fail to support the simultaneous acquisition of hyperspectral and full 3D spatial information.

In this work, we demonstrate a unified platform that integrates parallel 3D time-of-flight measurements with broadband infrared spectral ghost imaging, enabling real-time hyperspectral 3D imaging using single-pixel detectors. Technically, the system supports pulse-resolved spectral reconstruction across the 1100–1250 nm range with a spectral resolution of 1.4 nm, while simultaneously acquiring 3D point-cloud images within a 16.5 ns repetition period. These capabilities highlight its strong potential for advanced LiDAR applications.

## Basic principles

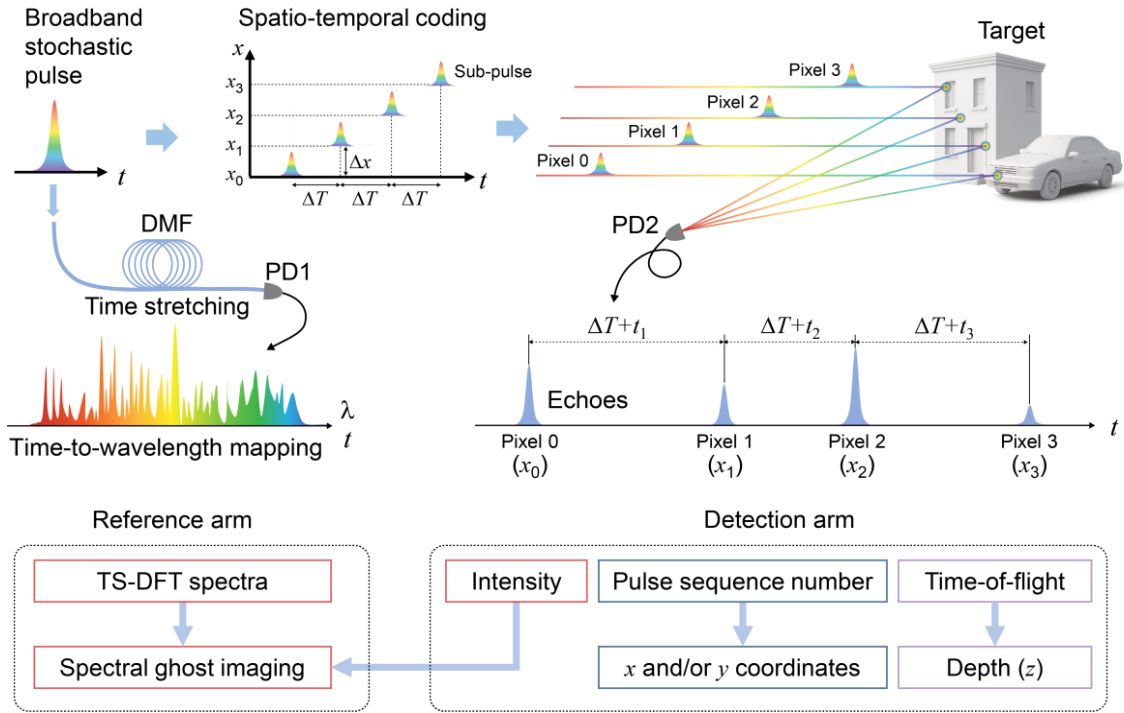

**Fig. 1 | Principle of operation.** Stochastic broadband pulses are split into a detection arm and a reference arm. In the detection arm, spatiotemporal encoding generates sub-pulse sequences (temporal spacing $\Delta T$; spatial interval $\Delta x$) that interrogate the target. The time-of-flight delay $t_i$ yields the depth $z_i$, which, together with the lateral coordinates $(x, y)$, reconstructs the 3D geometry. Each sub-pulse corresponds to an individual imaging pixel (indexed by $i$). In the reference arm, time-stretch dispersive Fourier transform (TS-DFT) performs time-to-wavelength mapping, and spectral ghost imaging retrieves the infrared

absorption spectrum by correlating the reflected pulses with the corresponding TS-DFT spectra. This unified framework enables hyperspectral 3D imaging across spatial ($x$, $y$, $z$) and spectral dimensions. Abbreviations: DMF, dispersion-managed fiber; PD, photodetector.

The working principle of the hyperspectral 3D LiDAR system is illustrated in Fig. 1. The system reconstructs the target's spectral response using TS-DFT-based spectral ghost imaging (SGI), while concurrently retrieving 3D spatial information via spatiotemporal encoding and ToF measurements. In this architecture, spectral and spatial information are encoded into the temporal waveform of each pulse and decoded in parallel, enabling unified, high-speed acquisition within a time-domain framework. To realize this, the system employs a stochastically modulated supercontinuum source, where each pulse carries a unique spectral profile that serves as a random encoding basis for SGI. The generated supercontinuum is subsequently split into a detection arm and a spectral reference arm for synchronized measurement and reconstruction.

For single-pulse sensing and imaging, the detection pulses are processed by a spatio-temporal encoding unit, which converts each pulse into an array of sub-pulses with fixed temporal spacing $\Delta T$ and spatial separation $\Delta x$. These sub-pulses sequentially illuminate the target, generating a train of echo signals. Each echo corresponds to a single imaging pixel: the lateral coordinate is given by $x_i = i \cdot \Delta x$, where $i$ denotes the sub-pulse index, while the axial coordinate is determined as $z_i = z_0 + c \cdot t_i / 2$, with $t_i$ representing the time of flight. The reference distance $z_0$, corresponding to the first sub-pulse, is obtained via temporal correlation between the detected echoes and the reference pulses [15, 33, 34].

The spectral response of the target, $R(\lambda)$, is reconstructed using a DFT-based SGI scheme. Specifically, the reconstruction is achieved by correlating the integrated intensities of the detection arm, $I_{\text{det}}$, with the spectrally resolved, time-stretched waveforms $I_{\text{ref}}(\lambda)$ measured in the reference arm. The normalized correlation function is defined as:

$$R(\lambda) = \frac{\left\langle \Delta I_{\text{ref}}(\lambda) \cdot \Delta I_{\text{det}} \right\rangle_N}{\sqrt{\left\langle \left[\Delta I_{\text{ref}}(\lambda)\right]^2 \right\rangle_N \left\langle \left[\Delta I_{\text{det}}\right]^2 \right\rangle_N}},$$

where $\langle \cdot \rangle_N$ denotes the ensemble average over $N$ measurements, and $\Delta I = I_i - I_{\text{mean}}$ represents intensity fluctuations relative to the ensemble mean.

Notably, this approach offers several key advantages. First, it enables simultaneous acquisition of spectral and 3D spatial information within a single framework. Second, it supports high-speed operation with single-shot capability at the pulse repetition rate. Third, the reconstruction relies solely on ToF analysis and second-order correlation, avoiding computationally intensive algorithms for high-dimensional data retrieval.

## Results

**Experimental setup**

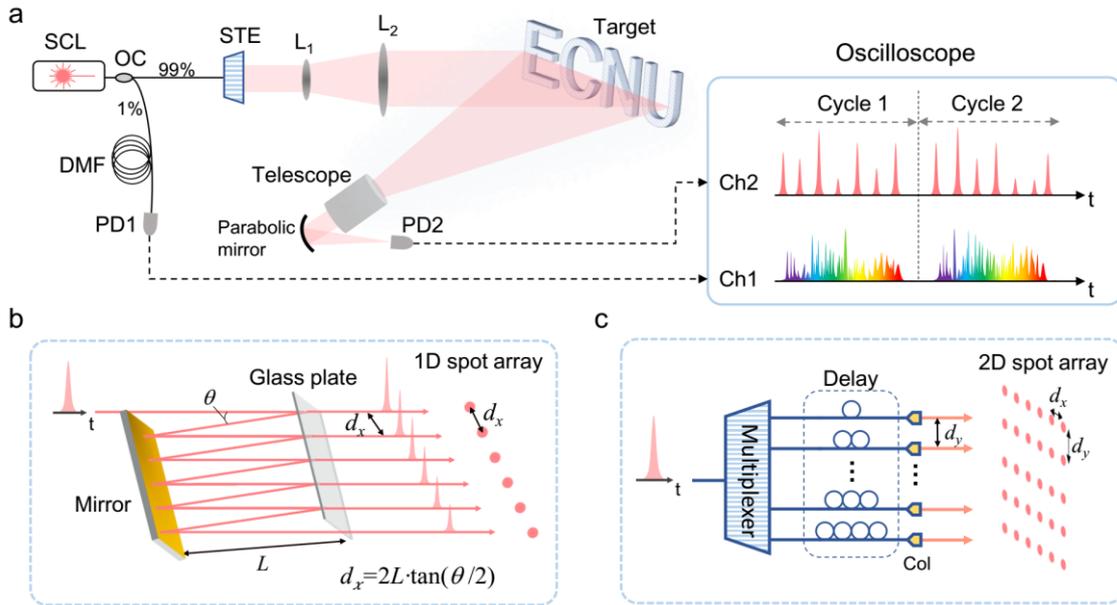

**Fig. 2 | Hyperspectral LiDAR system. a,** Experimental setup. SCL, supercontinuum laser; OC, optical coupler; L, lens; Col, collimator; STE, spatiotemporal encoder; PD, photodetector; DMF, dispersion-managed fiber. **b,** Schematic of STE, consisting of a high-reflectivity mirror and a thin glass plate. Multiple internal reflections generate a one-dimensional (1D) spot array. The incident angle $\theta$ is adjusted to 1.4°, yielding a spot spacing of $d_x$=1.8 mm for $L$=75 mm. **c,** Configuration for generating a two-dimensional (2D) spot array. A 1×16 optical multiplexer with length-differentiated fibers is integrated with the STE to form a 6×16 spot array. A pair of lenses (focal lengths 3 cm and 10 cm) expands the beam array, extending the vertical field of view to 6 cm.

As shown in Fig. 2a, the system utilizes a broadband laser that emits a train of spectrally unstable pulses at a repetition rate of 60.5 MHz (see Supplementary Fig. 1 for details). The source delivers an average output power of 150 mW over a spectral range of 1100-1250 nm. The laser output is split by a fiber coupler into a 1% reference arm and a 99% detection arm. In the reference arm, chromatic dispersion in a 500 m dispersion-managed fiber maps the pulse spectrum into the temporal domain. The resulting time-stretched waveform is detected by a 20 GHz photodetector and recorded using a high-speed oscilloscope, enabling single-shot spectral characterization.

In the detection arm, the beam is directed into a spatio-temporal encoder (STE), which generates an array of spatially separated and temporally delayed sub-pulses. For this proof-of-concept implementation, the STE consists of a high-reflectivity mirror and a thin glass plate (Fig. 2b). Multiple internal reflections produce a one-dimensional array of ~30 spots with uniform spatial spacing and fixed temporal delays. The temporal interval between adjacent sub-pulses, $\Delta T$, is determined by the mirror–plate separation $L$ as $\Delta T=2L/c$. With $L= 7.5$ cm, a temporal spacing of 500 ps is obtained. The lateral spacing $d_x$ between adjacent spots is given by $d_x =2L \cdot \tan(\theta/2)$, where $\theta$ is the incident angle. By incorporating a galvanometric scanner along the y-axis, the one-dimensional array can be extended to two dimensions. Alternatively, integrating a 1×16 optical multiplexer prior to the STE (Fig. 2c) enables direct generation of a two-dimensional spot array, allowing scan-less 3D imaging.

Following the STE, the beams are collimated by a two-lens system and used to illuminate targets forming the "ECNU" pattern. Each letter is fabricated from acrylic with a reflectivity below 10% and mounted on a translation stage to create a 3D scene. To demonstrate spectral resolving capability, 1 mm-thick cuvettes containing different chemical samples are placed in front of the corresponding letters. The transmitted and subsequently reflected light is collected by a telescope and focused onto a free-space, high-speed photodetector, whose output is recorded by the same oscilloscope.

## Systematic characterization

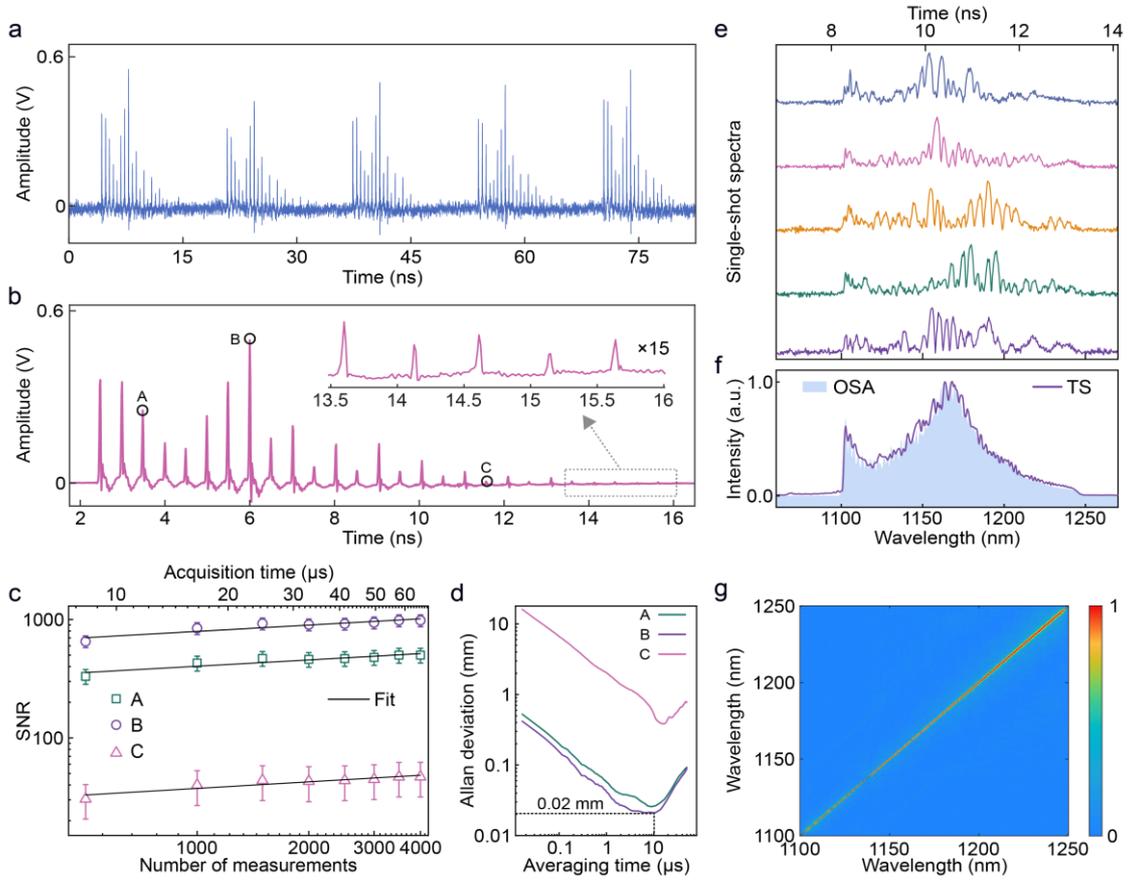

**Fig. 3 | Spatial and spectral characterization. a,** Pulse sequences over five consecutive cycles recorded by the oscilloscope, where each peak corresponds to the echo from an individual pixel. **b,** Averaged echo signals over 6044 cycles, with pulses at positions A, B and C selected for further analysis. **c,** Signal-to-noise ratio (SNR) at positions A, B and C as a function of acquisition time. **d,** Ranging precision at positions A, B and C as a function of averaging time; the Allan deviation reaches 0.02 mm at an averaging time of 10.4 μs. **e,** Five representative single-shot spectra recorded using the time-stretch dispersive Fourier transform (TS-DFT) technique. **f,** Comparison between TS-DFT spectra and a steady-state spectrum measured with an optical spectrum analyzer (OSA). **g,** Pearson correlation matrix of the filtered broadband spectrum, computed from an ensemble of 5000 single-shot TS-DFT measurements.

First, we characterize the system for spatial and spectral measurements. Figure 3a shows the recorded echo pulses over five consecutive cycles, each spanning 16.5 ns, corresponding to the initial pulse repetition period. The intensity variations of sub-pulses within a cycle arise from differences in their beam paths, which affect the coupling efficiency at the detector. In contrast, the variations of the same sub-pulse across cycles reflect the stochastic fluctuations of the initial laser pulses. In a single-shot measurement, the highest signal-to-noise ratio (SNR), defined as the ratio of the signal peak to the baseline noise standard deviation, is approximately 30, which can be enhanced by averaging. For example, Figure 3b presents the echo signals within a single cycle after 6044 averages (equivalent to an acquisition time of 100 μs), capturing up to 30 lateral pixels with a maximum SNR of 2300. Figure 3c shows the SNRs of different sub-pulses as a function of the number of averages or acquisition time. In the log-log plot, the fitted slopes are approximately 0.2, indicating that the SNR improvement deviates from the classical $\sqrt{N}$ scaling, which we attribute to the intrinsic laser instability. By calculating the flight time ($t_i$) of each sub-pulse, we extract their axial distances. The Allan deviation results in Fig. 3d demonstrate a highest ranging precision of 0.02 mm for the most intense peak at a cumulative averaging time of 10.4 μs.

Simultaneously, in the reference arm, we record TS-DFT spectra on a pulse-by-pulse basis (Fig. 3e). For time-to-wavelength calibration (Fig. 3f), 6044-fold averaged TS-DFT data are compared with a spectrum measured by an optical spectrum analyzer (OSA; Yokogawa AQ6375, 0.1 nm resolution). Limited by the 20 GHz detector bandwidth, the TS-DFT spectral resolution is 1.4 nm. The calibrated time-to-wavelength conversion factor is 34.6 ps/nm.

To assess spectral randomness, we computed the wavelength-to-wavelength Pearson correlation matrix from 5000 single-shot spectra (Fig. 3g). All off-diagonal correlations are below 0.05, confirming the mutual independence of spectral components—a key requirement for spectral-domain ghost imaging [29-31]. These highly uncorrelated patterns provide abundant independent information, enabling high-fidelity spectral reconstruction.

## Time-of-flight range imaging

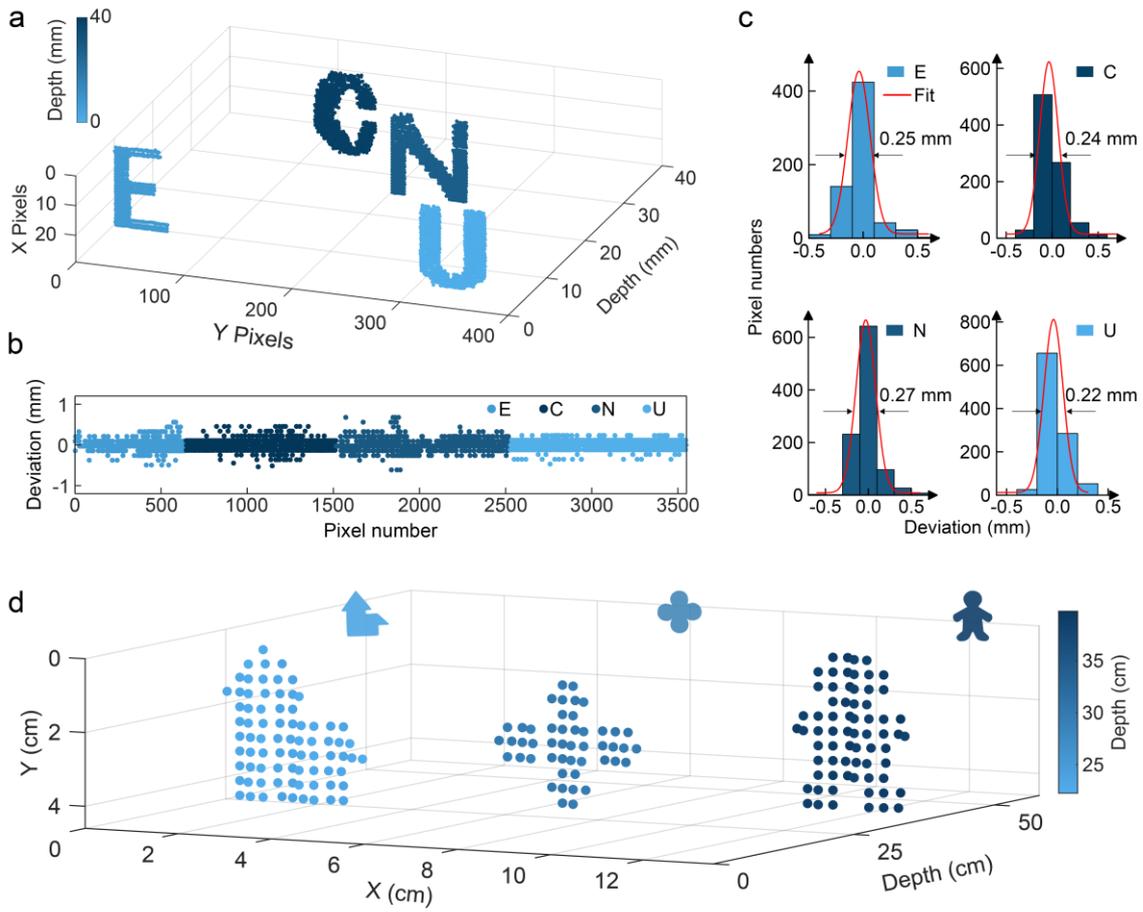

**Fig. 4 | Results of 3D ranging and imaging. a,** Reconstructed 3D point clouds of the objects ("E", "C", "N", and "U") positioned at different depths. **b,** Pixel-wise ranging deviations from the mean value for each object. **c,** Histograms of the ranging deviations in **b**, with red curves indicating Gaussian fits to the experimental data. The distance from the laser output to the reference plane (depth: 0 mm) is approximately 30 cm. **d**, 3D point clouds acquired using a two-dimensional (2D) spot array.

We next demonstrate the capability of our system for line-scanning 3D imaging. The 1D spot array is employed to perform vertical line scans across each letter target. The collected depth data from successive scans are aggregated to reconstruct the 3D scene, as shown in Fig. 4a. The maximum line-scan rate of the system is set by the laser repetition rate (i.e., 60.5 MHz), with approximately 30 pixels per line, corresponding to a point acuisition rate (PAR) of 1.8 GHz. With $\Delta T = 500$ ps, the non-ambiguous range (NAR) for a single-shot depth retieval is 7.5 cm. However, the determination of the reference distance ($z_0$) is not constrained by this ambiguity. This is because $z_0$ is obtained from intensity correlation between the detection and reference arms, rather than from ToF measurement of a pulse, and is further enabled by the use of a random laser (see details in Supplementary Fig. 2).

To achieve high ranging precision, 6044 measurements within a 100 μs acquisition window are accumulated for each line scan, resulting in a line-scan rate of 10 kHz. The deviation distributions, shown in Fig. 4b, are obtained by comparing the depth values of individual pixels with the corresponding mean value for each target. Figure 4c presents histograms of the depth deviations for all pixels across the four targets, each fitted with a Gaussian function. The extracted full width at half maximum (FWHM) values are 0.25, 0.24, 0.27, and 0.22 mm, respectively. The ranging precision is expected to improve with higher optical power or more sensitive detection (Supplementary Fig. 3).

Moreover, extending the spot array from 1D to 2D could enable single-snapshot 3D range imaging. As a demonstration, we place a 1×16 fiber multiplexer (TDS1315HA, Thorlabs) before the STE. The multiplexer channels have different fiber lengths, with a path difference of 36.8 cm between adjacent channels, corresponding to a delay of approximately 1.8 ns. This delay, combined with the STE's 1×6 spot generation, yields a 6×16 two-dimensional spot array (96 pixels in total), leading to a field of view of 2.7 cm in the lateral direction and 6 cm in the vertical direction (after beam expansion). To capture all pixel signals within a single repetition period, we reduce the laser repetition rate to 34.7 MHz (period: 28.8 ns) and set $\Delta T$ to 300 ps, resulting in a NAR of 4.5 cm at a PAR of 3.3 GHz. The imaging results are shown in Fig. 4d. Three polystyrene targets of different shapes are measured, and two interlaced scans are performed on each target to acquire the full images. The integration time per image is 980 ns. We note that the number of spots can be further increased, but doing so requires higher laser power and comes at the expense of a reduced repetition rate or a lower PAR. A detailed discussion of these trade-offs is provided in Supplementary Note 1.

## Hyperspectral 3D imaging

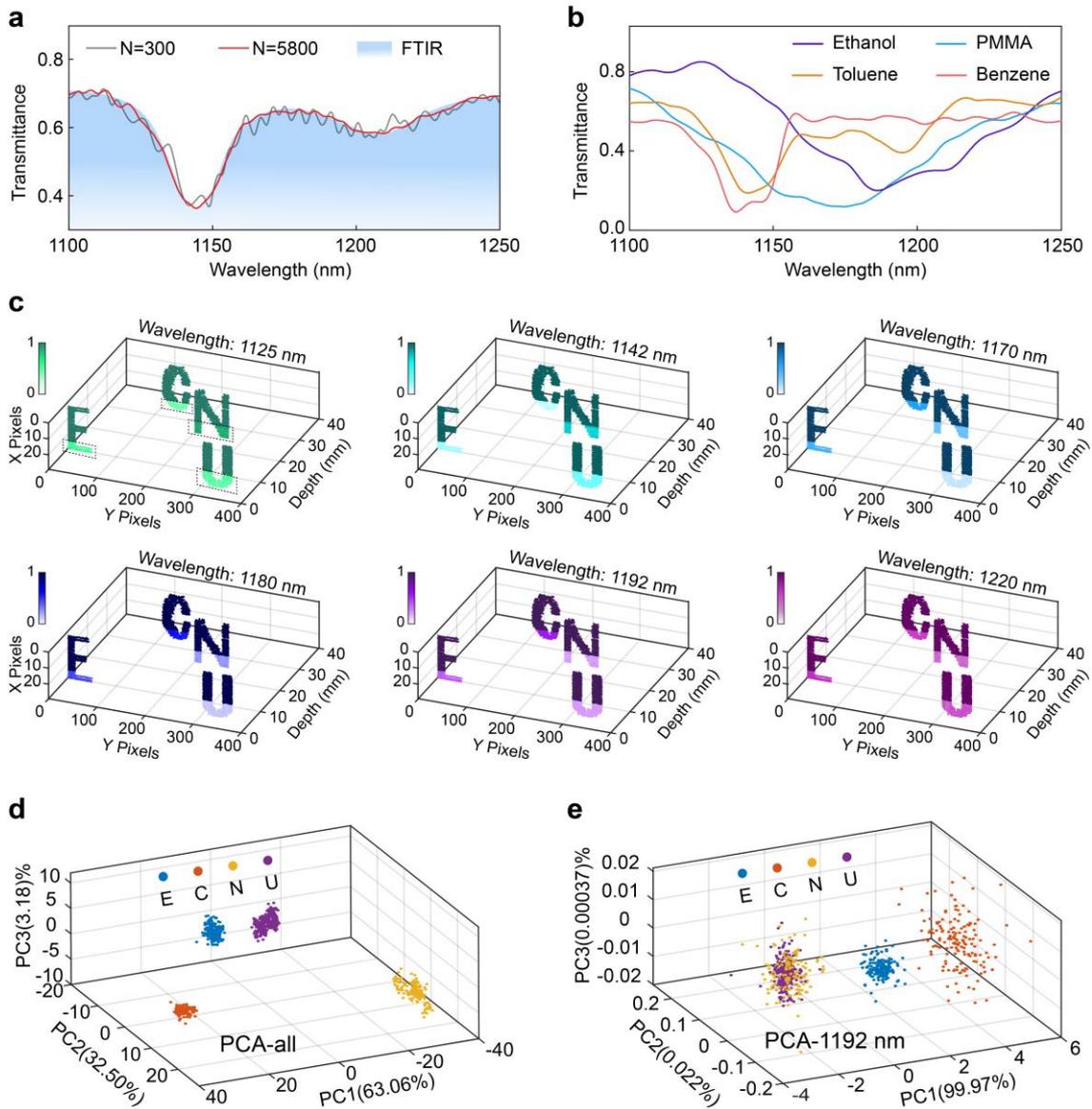

**Fig. 5 | Hyperspectral 3D imaging results. a,** Retrieved spectral-domain ghost imaging results for N = 300 and 5800 realizations, compared with spectra measured using a Fourier-transform infrared (FTIR) spectrometer with a resolution of 0.2 nm. **b,** Spectral ghost images of four samples (toluene, benzene, ethanol and PMMA) reconstructed from 5800 realizations. **c,** Monochromatic 3D images at selected wavelengths. **d, e,** Score maps in the 1120-1220 nm range (107 effective spectral elements with 1.4 nm resolution) and at 1192 nm, obtained via principal component analysis (PCA).

A key distinction of our system, compared with conventional 3D imaging techniques, is its ability to acquire spatial and spectral information simultaneously. For spectroscopic validation, we measure the transmission spectrum of a thin polystyrene film. Figure 5a compares the spectrum retrieved by SGI with that obtained using a commercial Fourier-transform infrared spectrometer (Invenio, Bruker), showing excellent agreement and thereby confirming the spectral fidelity of our approach.

We then extend 3D ToF ranging to hyperspectral imaging. As a demonstration, four chemical samples—benzene, ethanol, toluene, and polymethyl methacrylate (PMMA)—are positioned in front of the "ECNU" letters within the uniformly illuminated region at the bottom of the field of view. The corresponding transmission spectra extracted from our system are presented in Fig. 5b. These spectra, spanning 1100–1250 nm, exhibit distinct spectral signatures. By leveraging pixel-resolved hyperspectral information, we reconstruct wavelength-specific 3D images. For example, Fig. 5c shows monochromatic images at six selected wavelengths, revealing contrast arising from differential absorption. However, such monochromatic images provide limited discriminability, particularly for samples with similar absorption at selected wavelengths—reflecting a general limitation of conventional 3D LiDAR in material classification.

Finally, to highlight the advantage of our system for chemical or material identification, we apply principal component analysis (PCA) [35], a method that compresses high-dimensional spectral data into a few orthogonal components capturing the dominant variance. Applying PCA to the full hyperspectral data cube yields a 3D score plot (Fig. 5d) defined by the first three principal components, which together account for 98.7% of the total variance, indicating that the reduced representation preserves nearly all relevant spectral information. In this feature space, the four samples form well-separated clusters, enabling unambiguous discrimination independent of their spatial locations. In contrast, PCA performed on a single spectral channel (e.g., 1192 nm) produces a highly overlapping distribution, indicating poor separability (Fig. 5e). This comparison highlights the critical role of broadband spectral information in achieving robust chemical classification.

## Discussion

In summary, we demonstrate a 4D hyperspectral ghost-imaging LiDAR that unifies spectral and 3D spatial sensing within a single high-speed time-domain framework. By integrating SGI with TOF measurement, our system encodes both spectral and depth information into direct temporal signals. This architecture overcomes the fundamental speed limitations of conventional hyperspectral LiDAR (imposed by mechanical scanning and spectrometer-based detection). Experimentally, the system achieves a line-scanning rate of 60.5 MHz and a PAR of up to 3.3 GHz. With data averaging, the system further achieves 3D ranging with a depth precision of 0.02 mm, while simultaneously enabling transmission spectroscopy across the 1100–1250 nm band with a spectral resolution of 1.4 nm—sufficient for characterizing condensed-phase samples.

We note that the high-speed acquisition is achieved at the expense of a reduced NAR, reflecting a fundamental trade-off in ToF measurements [36, 37]. In addition, the current implementation is limited in pixel number, SNR, and detection sensitivity. These constraints, however, can be mitigated through several approaches, such as increasing the optical power via high-power fiber amplification and scaling the number of multiplexing.

Beyond these improvements, the proposed methodology provides a flexible and extensible framework in both detection modality and spectral coverage. For instance, integration with single-photon detection technologies [38] is expected to push operation into photon-limited regimes, enabling hyperspectral LiDAR under low-light conditions. Such capability is particularly relevant for reconstructing and spectrally characterizing surfaces with low reflectivity, high roughness, or low optical damage thresholds, including in biomedical contexts. Meanwhile, as demonstrated in prior studies [39-41], it can be adapted to frequency upconversion-based mid-infrared ghost imaging, thereby extending operation into the molecular fingerprint region (e.g., 3–5 μm). Besides, incorporating compressive sensing strategies [42-44] may provide a pathway to reduce data volume and accelerate spectral reconstruction.

By combining spectral discrimination with high-speed 3D imaging, our approach opens new opportunities for advanced remote sensing. Potential applications include environmental monitoring [45], where spectrally resolved LiDAR can identify pollutants while mapping their spatial distribution; precision agriculture [46], enabling simultaneous assessment of vegetation type, health, and morphology; and industrial process monitoring [41], where rapid tracking of 3D structure and chemical composition can support closed-loop control. In addition, integration with photonic multiplexing technologies may reduce system complexity and footprint [48], facilitating deployment in compact platforms such as autonomous navigation systems operating in complex environments.

# Methods

**Stochastic supercontinuum laser**

A home-built, all-polarization-maintaining linear-cavity ytterbium-doped fiber laser is used as the seed source. The 1.7 m cavity consists of a SESAM mode-locking mirror, a 0.6 m length of ytterbium-doped fiber (PM-YSF-HI-HP, Nufern) serving as the gain medium, and a chirped fiber Bragg grating (~70% reflectivity) for intracavity dispersion compensation and output coupling. Self-starting passive mode-locking is achieved at a pump power of 70 mW, generating pulses centered at approximately 1035 nm with a 3 dB bandwidth of 15 nm, a pulse duration of 1.1 ps, and a repetition rate of 60.5 MHz. The oscillator output (8 mW) is amplified to 2 W through a two-stage ytterbium-doped fiber amplifier, then compressed to ~190 fs using a transmission grating pair (PCG-1250/1030-986, Ibsen Photonics). The compressed pulses are launched into a 15 cm photonic crystal fiber (SC-3.7-975, NKT Photonics) to generate a supercontinuum spanning 800–1400 nm with an output power exceeding 200 mW. The supercontinuum is subsequently filtered to the 1100–1250 nm range using a 1100 nm long-pass and a 1250 nm short-pass filter, as this spectral region exhibits the most pronounced stochastic behavior arising from modulation instability.

**Light detection and data acquisition**

Returned echoes are collected by a telescope and detected using a single-pixel detector (EOT, ET3600; bandwidth >22 GHz). The detector output is recorded by a high-speed oscilloscope (DSOV334A, Keysight; 33 GHz bandwidth) for ToF measurements. For spectral ghost imaging, 1% of the supercontinuum is tapped off into a reference arm and transmitted through a 500 m dispersion-managed fiber. The resulting time-stretched reference signal is detected by a 20 GHz photodiode (KG-PD-20G, CONQUER) and recorded on a separate channel of the same oscilloscope. Spectral reconstruction is performed by correlating the integrated intensities of the returned pulses from the detection arm with the corresponding time-stretched reference spectra.

# References

1. Kim, I. et al. Nanophotonics for light detection and ranging technology. *Nat. Nanotechnol.* 16, 508–524 (2021).
2. Li, N. et al. A progress review on solid-state LiDAR and nanophotonics-based LiDAR sensors. *Laser Photonics Rev.* 16, 2100511 (2022).
3. Coddington, I., Swann, W. C., Nenadovic, L. & Newbury, N. R. Rapid and precise absolute distance measurements at long range. *Nat. Photon.* 3, 351–356 (2009).
4. Lee, J., Kim, Y.-J., Lee, K., Lee, S. & Kim, S.-W. Time-of-flight measurement with femtosecond light pulses. *Nat. Photon.* 4, 716–720 (2010).
5. Caldwell, E. D., Sinclair, L. C., Newbury, N. R. & Deschenes, J.-D. The time-programmable frequency comb and its use in quantum-limited ranging. *Nature* 610, 667–673 (2022).
6. Suh, M.-G. & Vahala, K. J. Soliton microcomb range measurement. *Science* 359, 884–887 (2018).
7. Wang, Z. et al. Rapid and precise distance measurement using balanced cross-correlation of a single frequency-modulated electro-optic comb. *Laser Photonics Rev.* 19, e01842 (2025).
8. Chang, B. et al. Dispersive Fourier transform based dual-comb ranging. *Nat. Commun.* 15, 4990 (2024).
9. Riemensberger, J. et al. Massively parallel coherent laser ranging using a soliton microcomb. *Nature* 581, 164–170 (2020).
10. Xu, W. et al. Progress and prospects for LiDAR-oriented optical phased arrays based on photonic integrated circuits. *npj Nanophoton.* 2, 14 (2025).
11. Li, B., Lin, Q. & Li, M. Frequency–angular resolving LiDAR using chip-scale acousto-optic beam steering. *Nature* 620, 316–322 (2023).
12. Zhang, X., Kwon, K., Henriksson, J., Luo, J. & Wu, M. C. A large-scale microelectromechanical-systems-based silicon photonics LiDAR. *Nature* 603, 253–258 (2022).
13. Jeong, D. et al. Spatio-spectral 4D coherent ranging using a flutter-wavelength-swept laser. *Nat. Commun.* 15, 1110 (2024).
14. Wang, Z. et al. High-precision time-domain stereoscopic imaging with a femtosecond electro-optic comb. *Nat. Commun.* 16, 6839 (2025).
15. Chen, R. et al. Breaking the temporal and frequency congestion of LiDAR by parallel chaos. *Nat. Photon.* 17, 306–314 (2023).
16. Gong, S. et al. Spectral-acoustic-coordinated astigmatic metalens for wide field-of-view and high spatiotemporal resolution 3D imaging. *Light Sci. Appl.* 15, 85 (2026).

17. Rogers, C. et al. A universal 3D imaging sensor on a silicon photonics platform. *Nature* **590**, 256–261 (2021).
18. Rivera, G., Porras, R., Florencia, R. & Sánchez-Solís, J. P. LiDAR applications in precision agriculture for cultivating crops: a review of recent advances. *Comput. Electron. Agric.* **207**, 107737 (2023).
19. Xing, C. et al. Fast-hyperspectral imaging remote sensing: emission quantification of $NO_2$ and $SO_2$ from marine vessels. *Light Sci. Appl.* **14**, 308 (2025).
20. Li, J., Yu, Z., Du, Z., Ji, Y. & Liu, C. Standoff chemical detection using laser absorption spectroscopy: a review. *Remote Sens.* **12**, 2771 (2020).
21. Chen, Y. et al. A 10-nm spectral resolution hyperspectral LiDAR system based on an acousto-optic tunable filter. *Sensors* **19**, 1620 (2019).
22. Ray, P., Salido-Monzú, D., Camenzind, S. L. & Wieser, A. Supercontinuum-based hyperspectral LiDAR for precision laser scanning. *Opt. Express* **31**, 33486–33499 (2023).
23. Li, D., Wu, J., Zhao, J., Xu, H. & Bian, L. SpectraTrack: megapixel, hundred-fps, and thousand-channel hyperspectral imaging. *Nat. Commun.* **15**, 9459 (2024).
24. Sun, B. et al. 3D computational imaging with single-pixel detectors. *Science* **340**, 844–847 (2013).
25. Xu, Y., Lu, L., Saragadam, V. & Kelly, K. F. A compressive hyperspectral video imaging system using a single-pixel detector. *Nat. Commun.* **15**, 1456 (2024).
26. Moreau, P. A., Toninelli, E., Gregory, T. & Padgett, M. J. Ghost imaging using optical correlations. *Laser Photonics Rev.* **12**, 1700143 (2018).
27. Ryczkowski, P., Barbier, M., Friberg, A. T., Dudley, J. M. & Genty, G. Ghost imaging in the time domain. *Nat. Photon.* **10**, 167–170 (2016).
28. Devaux, F., Moreau, P.-A., Denis, S. & Lantz, E. Computational temporal ghost imaging. *Optica* **3**, 698–701 (2016).
29. Rabi, S. et al. Spectral ghost imaging for ultrafast spectroscopy. *IEEE Photonics J.* **14**, 1–4 (2022).
30. Hu, J. et al. Broadband coherent Raman spectroscopy based on single-pulse spectral-domain ghost imaging. *Opt. Lett.* **50**, 6201–6204 (2025).
31. Amiot, C., Ryczkowski, P., Friberg, A. T., Dudley, J. M. & Genty, G. Supercontinuum spectral-domain ghost imaging. *Opt. Lett.* 43, 5025–5028 (2018).
32. Jiang, Y., Karpf, S. & Jalali, B. Time-stretch LiDAR as a spectrally scanned time-of-flight ranging camera. *Nat. Photon.* **14**, 14–18 (2020).
33. Zang, Z. et al. Ultrafast parallel single-pixel LiDAR with all-optical spectro-temporal encoding. *APL Photonics* **7**, 046102 (2022).

34. Zhang, X. et al. Spectrally encoded parallel LiDAR driven by super-bunching light. *Nat. Commun.* 17, 1161 (2026).

35. Greenacre, M. et al. Principal component analysis. *Nat. Rev. Methods Primers* 2, 100 (2022).

36. Qi, Y. et al. 1.79-GHz acquisition rate absolute distance measurement with lithium niobate electro-optic comb. *Nat. Commun.* **16**, 2889 (2025).

37. Li, R. et al. Ultra-rapid dual-comb ranging with an extended non-ambiguity range. *Opt. Lett.* **47**, 5309–5312 (2022).

38. Ren, X. et al. Single-photon counting laser ranging with optical frequency combs. *IEEE Photonics Technol. Lett.* **33**, 27–30 (2021).

39. Wen, Z. et al. Broadband up-conversion mid-infrared time-stretch spectroscopy. *Laser Photonics Rev.* **18**, 2300630 (2024).

40. Zhang, W. et al. Mid-infrared single-photon computational temporal ghost imaging. *Laser Photon. Rev.* **19**, 2402180 (2025).

41. Li, M. et al. Broadband terahertz comb with sub-Hz comb linewidth. *Adv. Photonics* **8**, 026015 (2026).

42. Barbastathis, G., Ozcan, A. & Situ, G. On the use of deep learning for computational imaging. *Optica* 6, 921–943 (2019).

43. Stellinga, D. et al. Time-of-flight 3D imaging through multimode optical fibers. *Science* 374, 1395–1399 (2021).

44. Li, M. et al. Neural-network-assisted spatio-spectral control of broadband light through multimode fibers. *Opt. Express* **33**, 54994–55004 (2025).

45. Ma, W. et al. Ozone pollution monitoring using a full-time hyperspectral tomography system for multiple air pollutants. *Nat. Commun.* **17**, 245 (2026).

46. Maeda, E. E. et al. Expanding forest research with terrestrial LiDAR technology. *Nat. Commun.* **16**, 8853 (2025).

47. Ameri, R., Hsu, C.-C. & Band, S. S. A systematic review of deep learning approaches for surface defect detection in industrial applications. *Eng. Appl. Artif. Intell.* **130**, 107717 (2024).

48. Chen, R. et al. Integrated bionic LiDAR for adaptive 4D machine vision. *Nat. Commun.* **17**, 24 (2026).